# Possible link of a structurally driven spin flip transition and the insulator-metal transition in the perovskite La$_{1-x}$Ba$_x$CoO$_3$

Peng Tong[1], Juan Yu,[1] Qingzhen Huang,[2] Kazuyoshi Yamada,[3] and Despina Louca[1*]

[1]*University of Virginia, Department of Physics, VA 22904, USA*
[2]*NIST Center for Neutron Research, Gaithersburg, MD 20899, USA*
[3]*WPI Advanced Institute for Materials Research,
Tohoku University, Katahira 2-1-1, Aoba, Sendai 980-8577, JAPAN*
(Dated: March 25, 2010)

The complex nature of the magnetic ground state in La$_{1-x}$A$_x$CoO$_3$ (A = Ca, Sr, Ba) has been investigated via neutron scattering. It was previously observed that ferromagnetic (FM) as well as antiferromagnetic (AFM) correlations can coexist prior to the insulator-metal transition (IMT). We focused on a unique region in the Ba phase diagram, from x = 0.17 - 0.22, in which a commensurate AFM phase appears first with a propagation vector, $k$ = (0, -0.5, 0.5), and the Co moment in the $(001)_R$ plane of the rhombohedral lattice. With increasing $x$, the AFM component weakens while an FM order appears with the FM Co moment directed along the $(001)_R$ (=$(111)_C$) axis. By $x$ = 0.22, a spin flip to new FM component appears as the crystal fully transforms to an orthorhombic ($Pnma$) structure, with the Co moments pointing along a new direction, $(001)_O$ (=$(110)_C$). It is the emergence of the magnetic $Pnma$ phase that leads to IMT.

The intricate coupling of the insulator to metal transition (IMT) with the magnetic state has captivated the physics of magnetoresistive oxides for a very long time. In the presumed metallic state, the interaction of the mobile charge with the spin should lead to a unique ground state that is typically ferromagnetic (FM) in nature. As the charge hoping enables the simultaneous coupling of two spins, the Double Exchange (DE) mechanism [1] wins over the superexchange interactions that favor the antiferromagnetic (AFM) ordering of the magnetic ions, the latter commonly observed in the parent phase. However, the ground state is more complex because it is a superposition of coexisting and competing magnetic states that yield emergent behaviors with very complex phases. The origin and organizations of such phases have been elusive because of the presence of more than one order parameter. At the core is the issue of magnetic phase separation that, in Mott systems in which strong correlations between the spin and the charge prevail, can lead to exotic states such as stripes as in superconductors [2] and colossal magnetoresistive materials [3]. Of relevance is the magneto-elastic coupling as well, but in systems such as the lanthanum cobaltite, LaCoO$_3$, this is less understood because of the lack of a lattice signature either via Jahn-Teller distortions, polaron formation or crystal transitions [4–7]. In this work, we show that the lattice degree of freedom plays a central role in the IMT via the percolation of magnetic droplets that become metallic after a complete structural transformation that brings upon a spin flip to a new magnetic order.

The parent compound, LaCoO$_3$, has a non-magnetic insulating ground state, with the Co$^{3+}$ ion in the low spin (LS) electronic configuration, $t_{2g}^6 e_g^0$ [8]. However, as the temperature rises, an electronic excitation occurs to a higher spin state, that fosters the development of dynamic FM and AFM correlations between the Co ions. This is consistent with the thermal excitation from the LS to the intermediate spin (IS) state. Coupled with this activation are single ion transitions of the Co$^{3+}$ within the IS state manifold, with a characteristic energy of 0.6 meV [9]. Upon doping La$_{1-x}$A$_x$CoO$_3$ by a divalent ion, it has been assumed that the percolation through the long-range FM insulating state yields to metallicity, presumed to develop around $x \sim 0.22$ in the case of A = Ba [7]. However, the true magnetic order and underlying physics are more complex than indicated from bulk property measurements. From our earlier elastic neutron scattering on single crystals of La$_{1-x}$A$_x$CoO$_3$ with A = Sr and Ba, we observed that in addition to the long-range FM order that appears with doping, an incommensurate magnetic (ICM) phase due to AFM spin correlations is also present, giving rise to an inhomogeneous ground state [10–12]. Only in La$_{1-x}$Ba$_x$CoO$_3$ does the ICM phase become commensurate when $x$ approaches 18 %. As both magnetic phases extend to long-range, their intensities become comparable but with different ordering temperatures [10]. Separate neutron powder diffraction studies on this system confirmed the presence of the long-range FM state, but failed to observe any signature of the AFM phase [13, 14].

The complexity of the magnetic ground state is intimately related to the crystal structure of La$_{1-x}$Ba$_x$CoO$_3$. From the neutron powder diffraction measurements in the range of 17 % $\leq x \leq$ 22% that encompasses the vicinity of the magnetic and transport transitions, we find that coexistence of the two magnetic orders thrives in the rhombohedral phase ($R\bar{3}c$) but when the lattice transforms to orthorhombic, only one phase remains. Initially

($x = 0.17$), a commensurate AFM state is present. With increasing $x$, a FM ordered phase appears and becomes dominant while the AFM component weakens. By $x = 0.22$, the AFM signal vanishes while a *new* FM structure develops. Coupled with the appearance of the new FM structure is a crystal symmetry transition, from the rhombohedral to an orthorhombic phase with $Pnma$ symmetry. Our result indicates that both AFM and FM components can spatially coexist, but the appearance of the orthorhombic phase is related to the onset of the IMT as it imposes a spin flip that creates metallic FM droplets. It is thus the percolation of these droplets that drives the system to become metallic. This would in turn suggest that the FM droplets in the rhombohedral phase are in fact insulating. This is fundamentally different from other magnetoresistive perovskites where only one FM transition is observed. The cobaltite system is unique as it presents an uncommon mechanism for the IMT, namely the change in the orbital overlap resulting from the expansion of the $ab$-plane of the orthorhombic symmetry brings about a new FM order that is conducive to hole hopping via the double exchange mechanism.

The samples were prepared by standard solid state reaction following the method in Ref. [11]. The neutron powder diffraction intensity data were collected using the BT-1 high resolution powder diffractometer at the NIST Center for Neutron Research. Data for $x = 0.17$, 0.19 and 0.22 were collected at 10 K using a wavelength of 1.5401 Å from a Cu (311) monochromator. Each data set was collected for more than 7.5 hours. Data for $x = 0.18$ were collected at 4 and 100 K for 18 hours each using a wavelength of 2.0784 Å from a Ge (311) monochromator. Weak $\lambda/3$ or $\lambda/2$ reflections are observed in the diffraction patterns and identified as indicated.

Fig. 1(a) is a plot of the low temperature neutron diffraction data for four compositions with the intensity plotted as a function of the momentum transfer, $Q$. The lattice coordinates for the rhombohedral symmetry of $x = 0.17$ - 0.19 are given in Table I. The rhombohedral angle, $\alpha$, in the primitive rhombohedron is also given to show how it gets smaller with increasing $x$. New diffraction peaks that could not be indexed by the nuclear symmetry were observed in $x = 0.17$. They can be described by a propagation vector of $k_1 = (0, -0.5, 0.5)$ determined from the use of the software $SARAh$ [15, 16] (Fig. 1(b)). This propagation vector corresponds to AFM order. Based on the nuclear symmetry and the observed magnetic propagation vector, the basis vectors ($BV$s) were calculated using $SARAh$ following the Irreducible Representation (IR) theory. For $k_1 = (0, -0.5, 0.5)$, there is only one nonzero representation, $\Gamma_1$, that has six associated $BV$s (following the KAREP notation [15]). The Co sites are separated into two groups with coordinates at Co1 (0, 0, 0) and Co2 (0, 0, 0.5). We find that the $BV$s $\psi_2$ (Co1:$m_x=0$, $m_y=1$, $m_z=0$; Co2: $m_x=-1$, $m_y=-1$, $m_z=0$) or $\psi_5$ ((Co1:$m_x=0$, $m_y=1$, $m_z=0$; Co2:

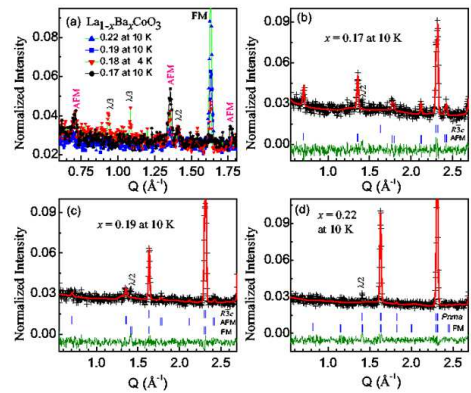

FIG. 1: (a) The neutron diffraction patterns in the low $Q$ range. The intensity is normalized to the highest nuclear Bragg peak. Diffraction peaks from the FM and AFM phases are marked. (b), (c) and (d) are the Rietveld refinement results for $x = 0.17$, 0.19 and 0.22, respectively. The observed intensities are plotted as symbols (cross) and the calculated patterns as solid lines. The $\lambda/3$ and/or $\lambda/2$ contaminations from the BT1 monochromator are marked.

$m_x=1$, $m_y=1$, $m_z=0$) alone can reproduce well the AFM diffraction pattern of $x = 0.17$. For $\psi_2$ (or $\psi_5$), the moments on Co1 and Co2 sites point along (0,1,0) and (-1,-1,0) (or (1,1,0)), respectively. The involvement of any other $BV$ does not improve the refinement. Note that in the powder diffraction, it is not possible to distinguish the moment orientation within the $ab$ plane of the $R\bar{3}c$ unit cell. Both $\psi_2$ and $\psi_5$ represent a non-collinear AFM configuration. The case of $\psi_5$ is shown in Fig. 2(a). The AFM unit cell is reproduced by expanding the nuclear unit cell by 2 $x$ $b$ and 2 $x$ $c$. Thus the AFM phase is commensurate with the nuclear unit cell, which is consistent with the data from single crystals [10].

As seen from Fig. 1(b), the nuclear (012) peak in the hexagonal setting which is equivalent to the (001) peak in cubic notation at $Q \sim 1.63$ Å$^{-1}$ is undetected due to its weak nuclear structure factor. At $x = 0.18$ (not shown in Fig. 1) and $x = 0.19$ (Fig. 1(c)) however, the intensity under this $Q$ position is remarkably enhanced, indicating the appearance of a FM component with a $k_2 = 0$. At the same time, the intensity at the AFM peak positions decreases as indicated through the change in the volume fractions given in Table 1. By $x = 0.19$, the AFM peaks diminish considerably, while the FM intensity becomes dominant. It has been previously shown that the FM intensity is isotropic due to the presence of FM droplets [9].

The coexistence of two magnetic orders in $x = 0.18$ and 0.19 makes the magnetic structure complex. Based on the nuclear structure refinement, there is no evidence for the presence of two nuclear symmetries that could give rise to two different magnetic states at these values of $x$. Fig. 3(a) is a plot of the fitting of the 10 K data of $x=0.19$ with the R$\bar{3}$c model. Thus, two options are possible, namely either the two magnetic orders originate from the same domain or from two different domains [17–19]. It is possible to combine the two phases if originating from a single magnetic domain and form a double wavevector magnetic structure. On the other hand, if independent because they originate from different magnetic domains, forming a single wavevector structure is more appropriate. The latter was chosen because as shown in Fig. 3(b) for $x = 0.18$, the strongest AFM intensity at $Q \backsim 1.36$ Å$^{-1}$ almost vanishes at 100 K, while the FM peak at $Q \backsim 1.63$ Å$^{-1}$ still persists at that temperature because of their different order parameters [10]. The FM order appears first with cooling, followed by the AFM order almost 40 K lower. In addition, from Fig. 1(a), although the ratio between the AFM to FM intensity changes significantly from $x = 0.18$ to 0.19, their propagation vectors do not change with $x$. From these observations we can deduce that the FM and AFM magnetic orders are weakly coupled to each other, namely they propagate in different domains [20–22]. To obtain the phase fraction of the two magnetic phases, we assumed that both magnetic orders have the same magnetic moment and the sum of their volume fractions is equivalent to that of the nuclear phase [18, 19]. In modeling the magnetic structure of $x = 0.18$, we used the AFM model of $x = 0.17$ as it works well in this case as well. In addition, for the $k_2 = 0$ propagation vector of the FM phase, there are three nonzero representations allowed by the symmetry. Of these, only $\psi_2$ (Co1:$m_x=0$, $m_y=0$, $m_z=1$; Co2: $m_x=0$, $m_y=0$, $m_z=1$) of the $\Gamma_3$ representation can reproduce the observed $k_2$ diffraction pattern alone. In this configuration for the FM structure, the moments for both Co1 (0,0,0) and Co2 (0,0,0.5) point along $(001)_R$ as shown in Fig. 2(b). The refined phase fractions of the AFM and FM phases are 48 % and 52 %, respectively. The refined moment is $1.33\mu_B$/Co. Thus by changing the concentration by 1 %, the AFM phase is suppressed while the FM phase develops and becomes as prominent as the AFM phase. By increasing the concentration further, at $x = 0.19$, the AFM fraction is reduced further to 17 % while the FM fraction increased to 83 %. The refined Co moment was determined to be 1.46 $\mu_B$. At the same time, since the AFM fraction is much less than the FM fraction, the increased moment in $x = 0.19$ originates from the enhanced FM coupling in the FM domains. Thus as $x$ increases, the FM domains grow in size and the FM coupling strengthens but the system is still insulating.

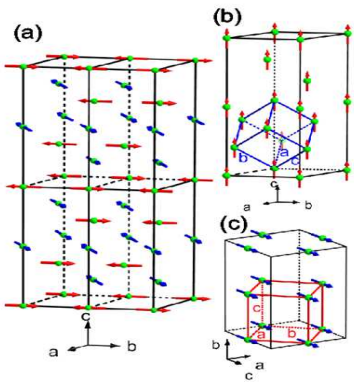

FIG. 2: Figure 2: (a) The AFM cell in the $R\bar{3}c$ phase. The moment is in the $(001)_R$ plane or in the $(111)_C$ plane; (b) The FM cell in the $R\bar{3}c$ phase. The moment is along the $(001)_R$ or $(111)_C$. (c) The new FM cell in the $Pnma$ phase. The moment is along the $(001)_O$ or $(110)_C$. This spin flip is associated with the crystal transition.

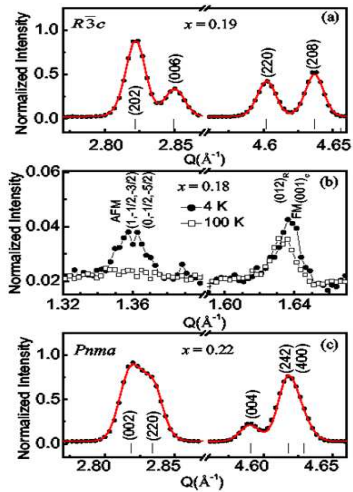

FIG. 3: (a) A comparison of the fit between the $x = 0.19$ with the $R\bar{3}c$ space group at 10 K. (b) The temperature dependence of the FM and AFM diffraction peaks for $x = 0.18$. The AFM peak disappears by 100 K while the FM peak remains. The FM order parameter sets in at ∼180 K as determined from single crystal data. (c) A comparison of the diffraction pattern of $x = 0.22$ at 10 K with the $Pnma$ model. Only when the structure fully transforms to this crystal phase that the new FM structure of Fig. 2(c) develops.

By $x = 0.22$, the AFM peaks are absent, while the peak intensity at $Q \backsim 1.63$ Å$^{-1}$ is enhanced further as seen in Fig. 1(d). However, the FM order appears with a new propagation vector of $k_3 = 0$. In addition, a crystal structure transition is observed at this composition where the nuclear phase changes to orthorhombic, with the space group $Pnma$ as shown in Fig. 3(c). The enhancement of the Bragg intensity at $Q \backsim 1.63$ Å$^{-1}$ should be attributed to the FM contribution rather than the $R\bar{3}c$-$Pnma$ transition, because the corresponding $Pnma$ nuclear structure factor is very weak at that point. It was previously reported that in $x = 0.20$, both the $R\bar{3}c$-$Pnma$ nuclear phases coexist at 4 K with a phase ratio of 48/50 [13]. However, in our $x = 0.22$ sample, the phase transition is almost complete and the residual rhombohedral phase is less than 5 % at 10 K. This suggests that the orthorhombic phase most likely coexists with the rhombohedral phase in a narrow region between $x = 0.2$ and 0.22. By $x = 0.22$, the AFM phase disappears and only an FM component with a new wavevector, $k_3$, is observed. For this propagation vector, there are four nonzero representations each with three $BV$s associated with the Co atoms on the 4b sites in the $Pnma$ lattice. The best refinement result can be obtained if the $BV$ $\psi_6$ (Co1:$m_x=0$, $m_y=0$, $m_z=2$; Co2: $m_x=0$, $m_y=0$, $m_z=2$; Co3:$m_x=0$, $m_y=0$, $m_z=2$; Co4: $m_x=0$, $m_y=0$, $m_z=2$) of $\Gamma_3$ is employed, in which the moments on all Co ions point to $(001)_O$ as shown in Fig. 2(c). The refined moment is 1.65 $\mu_B$/Co.

In the pseudo-cubic notation, the moment direction is along the face diagonal $(110)_C$ as shown inside of Fig. 2(c). For comparison, in the FM phase with the rhombohedral symmetry, the moment is along the body diagonal $(111)_C$ as shown in Fig. 2(b), while the AFM moment is in the $(111)_C$ plane. In the $R\bar{3}c$ lattice, the Co atoms align on the $\bar{3}$ axis. In the FM phase, an inversion operation needs to be added to keep the spin direction under the symmetry operation [23]. The operation of the threefold axis requires the moment point long $(001)_R$ because any other spin orientation cannot be conserved. In the $Pnma$ lattice, on the other hand, the Co atoms are located on the $2_1$ screw axis either along $(100)_O$, or $(010)_O$, or $(001)_O$. Under the $2_1$ operation along $(100)_O$, the spin can only point to $(100)_O$, or perpendicular to it with the addition of the inversion operation. If it is perpendicular to the $(100)_O$ axis, the spin must point to $(010)_O$ or $(001)_O$ otherwise its orientation cannot be kept under the operations of the $2_1$ axis. As a result, in the FM state of the $Pnma$ lattice, the Co spin may point either along $(100)_O$, $(010)_O$, or $(001)_O$.

What then happens to the AFM domains? As the distance between the AFM planes increases with increasing Ba content, the AFM spin direction continuously changes from the $(111)_C$ plane to the $(111)_C$ axis. This is supported by the continuous decrease of the volume fraction of this component, leading to the weakening of the AFM phase by $x = 0.20$ [13]. Thus the FM moment for $x \leq 0.20$ points along the body diagonal of the pseudocube. This FM phase is however insulating where the Co-O-Co bonds are buckled with an angle of 167.7$^o$, which reduces the orbital overlap. With the complete transformation to the *orthorhombic* phase, only one magnetic component is observed, and in the pseudocubic notation, the moment points along the $(110)_C$ face diagonal. In this symmetry, the Co-O-Co bond angle between pairs in the $ab$-plane increases to 170.5° while the bond length increases from 3.845 to 3.855 Å as well. With this lattice distortion, all bond pairs become FM coupled as the distance between the $(111)_C$ planes increases. This maximizes the Co 3d orbital overlap that not only does it allow for the spins to couple ferromagnetically but also allow the hopping of charges. To summarize, the coexistence of the AFM and FM orders as well as the $R\bar{3}c$-$Pnma$ transition add new insights to the IMT mechanism in La$_{1-x}$Ba$_x$CoO$_3$. The nucleation of the orthorhombic phase in the rhombohedral matrix predisposes the lattice to the new magnetic order. In a narrow region of the phase diagram prior to the IMT, the orthorhombic domains grow quickly, but it is only the formation of the new FM order that that enables the percolation of charge.

This work is supported by the U. S. Department of Energy under contract DE-FG02-01ER45927 and the U. S. DOC through NIST-70NANBH1152.

Table I. The refinement results of the nuclear and magnetic structures of La$_{1-x}$Ba$_x$CoO$_3$. In $x = 0.17$- 0.19, the symmetry is R$\bar{3}$c (in hexagonal setting) with two equivalent Co ions at (000) and (00$\frac{1}{2}$) and lattice constants, a = 5.4593(2) Å and c = 13.2062(1) Å in $x = 0.17$, a = 5.4587(4) Å and c = 13.2186(2) Å in $x = 0.18$, and a = 5.4611(3) Å and c = 13.2289(1) Å in $x = 0.19$. In $x = 0.22$, the symmetry is $Pnma$ with four equivalent Co ions at (00$\frac{1}{2}$), ($\frac{1}{2}\frac{1}{2}$0), (0$\frac{1}{2}\frac{1}{2}$) and ($\frac{1}{2}$00) and a = 5.4309(2) Å, b = 7.6835(6) Å and c = 5.4765(2) Å. The rhombohedral angle, $\alpha_R$, Co-O-Co angle ($\beta$), phase fraction ($f$) in %, the moment per Co site ($\mu$) and the spin orientation (SO) in the FM and AFM structures are listed.

| x | 0.17 | 0.18 | 0.19 | 0.22 |
|---|---|---|---|---|
| $\alpha_R^o$ | 60.55(1) | 60.51(2) | 60.49(1) | - |
| $\beta^o$ | 167.13(3) | 167.66(8) | 167.77(3) | 165.89(7)/170.49(4) |
| $f_{AFM}$ | 100 | 48 | 17 | 0 |
| $f_{FM}$ | 0 | 52 | 83 | 100 |
| $\mu(\mu_B)$ | 1.20(2) | 1.33(1) | 1.46(2) | 1.65(7) |
| AFM SO | Co$_1$: $(010)_R$; Co$_2$: $\pm(110)_R$ | | | No AFM |
| FM SO | No FM | Co$_{1,2}$: $(001)_R$ | | Co: $(001)_O$ |

*To whom correspondence should be addressed.

---